# Graph Data Models and Relational Database Technology

Malcolm Crowe
Emeritus Professor, Computing Science
University of the West of Scotland
Paisley, United Kingdom
Email: Malcolm.Crowe@uws.ac.uk

Fritz Laux
Emeritus Professor, Business Computing
Universität Reutlingen
Reutlingen, Germany
Email: Fritz.Laux@reutlingen-university.de

*Abstract*—Recent work on database application development platforms has sought to include a declarative formulation of a conceptual data model in the application code, using annotations or attributes. Some recent work has used metadata to include the details of such formulations in the physical database, and this approach brings significant advantages in that the model can be enforced across a range of applications for a single database. In previous work, we have discussed the advantages for enterprise integration of typed graph data models (TGM), which can play a similar role in graphical databases, leveraging the existing support for the unified modelling language UML. Ideally, the integration of systems designed with different models, for example, graphical and relational database, should also be supported. In this work, we implement this approach, using metadata in a relational database management system (DBMS).

## I. INTRODUCTION

For many years, the process of database implementation has included a conceptual data modeling phase, and this has often been supported by declarative structures using annotations or attributes [1]. Some recent DBMS have included metadata in the relational model to form a bridge with the physical database. This approach brings significant advantages in that the data model can be enforced across all applications for a single database. In previous work [2], we provided mapping rules for TGM so that data models can play a similar role in graphical databases, using the notations of UML [3]. During such early conceptual model building, incremental and interactive exploration can be helpful [4] as fully automated integration tools may combine things in an inappropriate way, and the use of data types [5] can help to ensure that semantic information is included not merely in the model, but also in the final database. In this short paper we report on such an implementation of TGM, using metadata in a relational DBMS [6], partly inspired by recent developments in the PostgreSQL community [7].

As with the original relational model, the Typed Graph Model (TGM) has a rigorous mathematical foundation as an instance of a Graph Schema.

The plan of this short paper is to review the TGM in Section II, and discuss the implementation details in Section III, including an illustrative example. Section IV provides some conclusions.

## II. THE TYPED GRAPH MODEL AND INFORMATION INTEGRATION

We will construct a TGM for a database by declaring instances of nodes and edges as an alternative to specifying tables of nodes and edges.

### A. Typed Graphs formalism

In this section we review the informal definition of the TGM from [2], using small letters for elements (nodes, edges, data types, etc.) and capital letters for sets of elements. Sets of sets are printed as bold capital letters. A typical example would be $n \in N \in \mathbf{N} \subseteq \wp(N)$, where $N$ is any set and $\wp(N)$ is the power-set of $N$.

Let $T$ denote a set of simple or structured (complex) data types. A data type $t:=(l,d) \in T$ has a name $l$ and a definition $d$. Examples of simple (predefined) types are ($int, \mathbb{Z}$), ($char, ASCII$), (%,[0..100]) etc. It is also possible to define complex data types like an order line ($OrderLine, (posNo, partNo, partDescription, quantity)$). The components need to be identified in $T$, e. g., ($posNo, int>0$). Recursion is allowed as long as the defined structure has a finite number of components.

**Definition 1 (Typed Graph Schema, TGS)** *A typed graph schema is a tuple TGS=($N_S, E_S, \varrho, T, \tau, C$)    where:*

- $N_S$ *is the set of named (labeled) objects (nodes) n with properties of data type $t:=(l,d) \in T$, where l is the label and d the data type definition.*

- $E_S$ *is the set of named (labeled) edges e with a structured property $p:=(l,d) \in T$, where l is the label and d the data type definition.*

- $\varrho$ *is a function that associates each edge e to a pair of object sets (O,A), i. e., $\varrho(e):=(O_e, A_e)$ with $O_e, A_e \in \wp(N_S)$. $O_e$ is called the* tail *and $A_e$ is called the* head *of an edge e.*

- $\tau$ *is a function that assigns for each node n of an edge e a pair of positive integers $(i_n, k_n)$, i. e., $\tau_e(n):=(i_n, k_n)$ with $i_n \in \mathbb{N}_0$ and $k_n \in \mathbb{N}$. The function $\tau$ defines the min-max multiplicity of an edge connection. If the min-value $i_n$ is 0 then the connection is optional.*

- *C is a set of integrity constraints, which the graph database must obey.*

The notation for defining data types T, which are used for node types $N_S$ and edge types $E_S$, can be freely chosen: and in this implementation SQL will be used for identifiers and expressions, together with a strongly typed relational database engine. The integrity constraints C restrict the model beyond the structural limitations of the multiplicity τ of edge connections. Typical constraints in C are semantic restrictions of the content of an instance graph.

**Definition 2 (Typed Graph Model)** *A typed graph Model is a tuple TGM=(N,E,TGS,φ) where:*

- *N is the set of named (labeled) nodes n with data types from $N_S$ of schema TGS.*
- *E is the set of named (labeled) edges e with properties of types from $E_S$ of schema TGS.*
- *TGS is a typed graph schema as defined above..*
- *φ is a homomorphism that maps each node n and edge e of TGM to the corresponding type element of TGS, formally:*

$$\varphi: TGM \rightarrow TGS$$
$$n \mapsto \varphi(n) := n_S (\in N_S)$$
$$e \mapsto \varphi(e) := e_S (\in E_S)$$

The fact that φ maps each element (node or edge) to exactly one data type implies that each element of the graph model has a well-defined data type. The homomorphism is structure preserving. This means that the cardinality of the edge types is enforced, too. In this implementation, the declaration of nodes and edge of the TGM develops the associated TGS incrementally including the development of the implied type system *T*. Data type and constraint checking is applied for all nodes and edges before any insert, update, or delete action can be committed.

*B. The Data Integration Process*

The full benefit of information integration requires the integration of source data with their full semantics. We believe a key success factor is to model the sources and target information as accurately as possible. The expressive power and flexibility of the TGM allows to describe the meta-data of the sources and target precisely and in the same model, which simplifies the matching and mapping of the sources to the target. The tasks of the data integration process are:

1) model sources as TGS $S_i$ ($i = 1, 2, ..., n$)
2) model target schema *T* as TGS *G*
3) match and map sources $S_i$ with TGS *G*
4) check and improve quality
5) convert TGS G back to T again

Steps 3 and 4 can occur together in an interactive process once the basic model has been outlined. Such a process is crucial for EII and other data integration projects, which demand highly accurate information quality, which can be further improved with the use of different mappings.

To start the process, it may be necessary to collect structure and type information from a data expert or from additional information. Where sources are databases, the rigid structures provide a good starting point. Otherwise, the relevant data must first be identified together with its meta-data if available. This includes coding and names for the data items. The measure units and other meta-data provided by the data owner are used to adjust all measures to the same scale. The paper of Laux [5] gives some examples how to transform relational, object oriented, and XML-schemata into a TGS.

If the source is unstructured or semi-structured, e.g., documents or XML/HTML data, concepts and mechanisms from Information Retrieval (IR) and statistical analysis may help to identify some implicit structure or identify outliers and other susceptible data. If the data are self-describing (JSON, key-value pairs, or XML) linguistic matching can be applied with additional help from a thesaurus or ontology. Nevertheless, it is advisable to validate the matching with instance data or an information expert. We present two possible TGS for a single enterprise in UML notation in Figure 1. This little example demonstrates already the flexibility of the model in terms of detail and abstraction.

### III. IMPLEMENTATION IN THE RELATIONAL DATABASE SCHEMA

The prerequisite for implementation of a typed graph modelling system is to have a strong type system in the RDBMS. If this is already available, then a graph modelling capability can be added relatively simply, with slight extensions to the normal SQL syntax for creating and altering structured types, and some metadata for distinguishing node and edge types from other kinds of structured types.

Then the main difference between a graph schema as described above and a schema in most DBMS is that columns and attributes of database tables have a predefined order. In addition, for a given node type or edge type, there is a single base table containing the instances of that type. One way to build a graph is to insert rows in these tables.

The aim of additional graph support in the DBMS is to simplify the tasks of graph definition and searching. We add CREATE and MATCH statements, which we describe next.

*A. Graph-oriented syntax added to SQL*

To the normal SQL CREATE syntax, we add an option for constructing a graph inline:

```
Create: CREATE Node { Edge Node } {',' Node {
Edge Node }}.
  Node: '(' NewG | id ')'.
  NewG: id {':' label } [doc] .
  –Edge: '-[' NewG ']->' | '<-[' NewG ']-' .
```

In this syntax, the strings enclosed in single quotes are tokens, including several new token types for the TGM. In corresponding source input, unquoted strings are used for case-insensitive identifiers and double quoted strings for case-sensitive identifiers, possibly containing other Unicode characters. As usual in SQL, string constants in input will be single quoted, and doc is a JSON-like structure providing a set of properties and value expressions, possibly including metadata definitions for ranges and multiplicity.

Such declarative statements build a base table in the database for each label.

Nodes and edges and new node types and edge types can be introduced with this syntax. The database engine constructs a base table for each distinct label, with columns sufficient to represent the associated properties. These database base tables for node types (or edge types) contain a single row for each node (resp. edge) including node references. They can be equipped with indexes, constraints, and triggers in the normal ways.

To the normal SQL DML, we add the syntax for the MATCH query, which has a similar syntax, except that it may contain unbound identifiers for nodes and edges, their labels and/or their properties.

```
Match: MATCH Node {',' Node } [WhereClause] Statement .
```

The first part of the MATCH clause defines a graph expression. We say that a graph expression is bound if it contains only constant values, and all its nodes and edges are in the database. The MATCH algorithm proceeds along the node expressions, matching more and more of its nodes and edges with those in the database by assigning values to the unbound identifiers. If we cannot progress to the next part of the MATCH clause, we backtrack by undoing the last binding and taking an alternative value. If the processing reaches the end of the MATCH statement, the set of bindings contributes a row in the default result, subject to the optional WHERE condition. These rows then act as a source of values for the following statement.

### B. Outline of the usage of the TGM

Following the suggestion in [5] we will consider the use of the TGM in analysis, where an interactive process is envisaged. The nodes and edges contained in the database combine to form a set of disjoint graphs that is initially empty. Adding a node to the database adds a new entry to this set. When an edge is added, either the two endpoints are in the same graph, or else the edge will connect two previously disjoint graphs. If each graph in the set is identified by a representative node (such as the one with the lowest uid) and maintains a list of the nodes and edges it contains, it is easy to manage the set of graphs as data is added to the database.

If an edge is removed, the graph containing it might now be in at most two pieces: the simplest algorithm removes it from the set and adds its nodes and edges back in.

The database with its added graph information can be used directly in ordinary database application processing, with the advantage of being able to perform graph-oriented querying and graph-oriented stored procedures. The normal processing of the database engine naturally enforces the type requirements of the model, and also enforces any constraints specified in graph-oriented metadata. The nodes and edges are rows in ordinary tables that can be accessed and refined using normal SQL statements. In particular, using the usual dotted syntax, properties can be SET and updated, and can be removed by being set to NULL.

As the TGM is developed and merged with other graphical data, conflicts will be detected and diagnostics will help to identify any obstacles to integrating a new part of the model, so that the model as developed to that point can be refined.

### C. An example

To get started with a customer-supplier ordering system we could have a number of problematic CREATE statements such as:

```
CREATE
   (Joe:Customer {"Name":'Joe Edwards',
Address:'10 Station Rd.'}),
   (Joe)-[:Ordered {"Date":date'22/11/2002'} ]->
(Ord201:"Order")-[:Item {Qty: 5}]->("16/50x100" :
WoodScrew),
   (Ord201)-[:Item {Qty: 5}]->("Fiber 12cm":
WallPlug),
   (Ord201)-[:Item {Qty: 1}]->("500ml" :
RubberGlue)
```

Primary keys for edges are here being left to the engine to supply – they could be specified explicitly if preferred. Name, Order and Date are in double quotes because they are reserved words in SQL. By default, the entire CREATE statement shown is considered a single transaction by the database engine: if the syntax checker is happy with it, it will be automatically committed.

It is easy to criticize what the user offers here: and the graph would benefit from splitting up composite information such as Fibre 12cm and 16/8x100 to clarify the meaning of the components and facilitate processing. Such changes can be made by the designer later.

Assuming the database is empty before we start, the first line above, if committed, would create a new base table CUSTOMER (a NodeType)

```
CREATE TYPE CUSTOMER as ("Name" char, ADDRESS char) NodeType
```

The NodeType metadata flag adds as the first column a primary key column ID of type char so that the new CUSTOMER table has an initial row

```
('JOE','Joe Edwards','10 Station Rd.')
```

That would work. The next line defines four more base tables, two NodeTypes and two EdgeTypes:

```
CREATE TYPE "Order" NodeType
CREATE TYPE WOODSCREW NodeType
CREATE TYPE ORDERED as ("Date" date)
    EdgeType(CUSTOMER,"Order")
CREATE TYPE ITEM as (QTY int)
    EdgeType("Order",WOODSCREW)
```

This also will work, but is probably not what the analyst wanted, because the Item edge type connects to nodes of type WOODSCREW. If this is committed, we cannot later have an Item edge connecting to a WALLPLUG.

But nothing is committed yet, so when the database engine finds this difficulty, it simply replaces the specification `:WoodScrew` in the second line by `:WoodScrew:&1` , and similar changes to WallPlug and RubberGlue.

This adds a new anonymous base node type for these node types, with a system-generated name

```
CREATE TYPE &1 NodeType
```

and the node type proposal becomes

```
CREATE TYPE ITEM as (QTY int)
    EdgeType("Order",WOODSCREW)

CREATE TYPE WoodScrew UNDER &1
CREATE TYPE WallPlug UNDER &1
CREATE TYPE RubberGlue UNDER &1
```

The analyst can be advised that this has been done, and they can later choose a better name for the new NodeType &1 (maybe PRODUCT?). This process of generalization can be offered as a standard database transformation.

After the nodes and edges have been generated and the transaction commits, the node and edge data would be installed in the database as follows:

```
CUSTOMER ('Joe','Joe Edwards',
    '10 Station Rd.')
"Order" ('Ord201')
WOODSCREW ('16/50x100')
WALLPLUG ('Fiber 12cm')
RUBBERGLUE ('500ml')
ORDERED ('&2','Joe','Ord201',
    'date'22/11/2002')
ITEM ('&3','Ord201','16/50x100')
    ('&4','Ord201','Fiber 12cm')
    ('&5','Ord201','500ml')
```

This is satisfyingly neat. We see that while the metadata flag NodeType gave the node type a primary key as the first column ID that is a primary key, the metadata flag EdgeType has given the edge types three initial columns: ID, a primary key, LEAVING, a foreign key to the leaving node type, and ARRIVING, a foreign key to the arriving type. Note also that ITEM's arriving type is the new anonymous type &1.

It is noteworthy that this mechanism allows schemas to evolve bottom-up during the database design, as envisaged in [2]. The normal Schema-first strategy is still available, and the two approaches can be combined for convenience. Either way, the database will contain a rigorous and enforceable relational schema at all stages, since any declarations that would not be enforceable will be rejected before being committed to the database.

During refinement of the model, there are opportunities for adding constraints and other metadata. Such details, and the enhanced diagnostics mentioned above, are the subject of ongoing research. The conference presentation will provide an opportunity for a demonstration of the process and more details on MATCH.

IV. CONCLUSIONS

The purpose of this paper was to report some progress in our Typed Graph Modeling workstream. The work is available on Github [8] for free download and use and is not covered by any patent or other restrictions.

The current "alpha" state of the software implements all of the above ideas apart but lacks the suggested interaction with the model designer. The test suite includes a version of the running example together with others that demonstrate the integration of the relational and typed graph model concepts in Pyrrho DBMS.

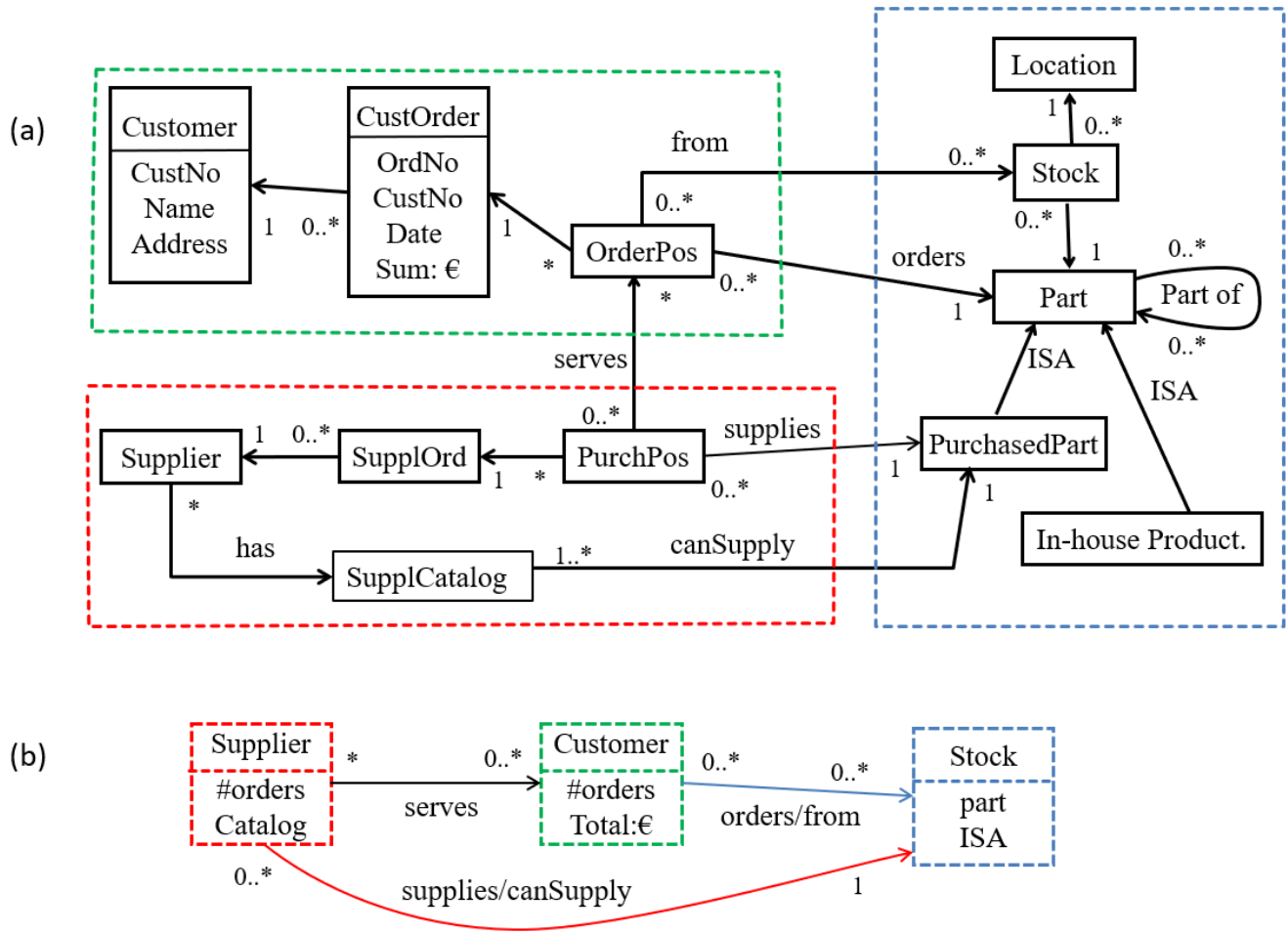

Figure 1. Example TGM of a commercial enterprise showing two levels of detail